\documentclass[aps,twocolumn,prb,showpacs,10pt,floatfix,groupedaddress]{revtex4-1}
\usepackage{amssymb}
\usepackage{amsmath}
\usepackage{graphicx}
\usepackage{braket}
\usepackage{dcolumn}
\usepackage{bm}
\usepackage{textcomp}
\usepackage{color}
\usepackage{braket}
\usepackage{amsthm}
\usepackage{url}
\usepackage{mathtools}

\begin{document}

\title{Topological edge states on time-periodically strained armchair graphene nanoribbons}

\author{Pedro Roman-Taboada}  
\email{pedro.roman.taboada@gmail.com}
\author{Gerardo G. Naumis}

\affiliation{Departamento de Sistemas Complejos, Instituto de
 F\'{i}sica, Universidad Nacional Aut\'{o}noma de M\'{e}xico (UNAM), 
Apartado Postal 20-364, 01000 M\'{e}xico, Ciudad de M\'{e}xico,
 M\'{e}xico}


\begin{abstract}
We report the emergence of electronic edge states in time-periodically driven strained armchair terminated graphene nanoribbons. This is done by considering a short-pulse spatial-periodic strain field. Then, the tight-binding Hamiltonian of the system is mapped into a one dimensional ladder. The time periodicity is considered within the Floquet formalism. Thus the quasienergy spectrum is found numerically by diagonalizing the evolution operator. For some particular cases, the quasienergy spectrum is found analytically. We find that the system is able to support gapless and gapped phases. Very different edge states emerge for both the gapless and the gapful phases. In the case of the gapped phase, edge states emerge at the gap centered at zero quasienergy, although the Chern number is zero due to the chiral symmetry of the system. For the gapless phase, besides edge states at zero quasienergy, cosine like edge states which merge and coexist with the bulk band are observed. To confirm the topological nature of these edge states, we analytically obtained the effective Hamiltonian and its spectrum for a particular case, finding that the edge states are topologically weak. Finally, we found analytically the evolution of band edges and its crossings as a function of the driven period. Topological modes arise at such crossings. 
\end{abstract}


\maketitle

\section{Introduction}
\label{intro}

Topological insulators are materials which can support topologically protected low energy excitations at the edges\cite{Qi11}. Such low energy excitations have attracted a lot of attention due to their potential to be used in the field of topological quantum computing\cite{tqc08,Ma17} or in spintronics\cite{spintronics04,SpinTakehito14,Majspin17}. After the experimental observation of topological insulators\cite{topMoore10,Chen178}, many systems exhibiting topologically non-trivial properties have been proposed\cite{Indu13,Avila2013,FloTop13,thakurathi13,Usaj14,KitaevDip14,Hossein15,Yuce2015,WeakTop15,SSHdriven15,klinovaja16,Sedlmayr16,Mitra16,Agarwala16,Amit16,Akzyanov16,Agarwala17,Roman17,jelena117}. Among them, one can mention the remarkable case of periodically driven systems, which have been proven to have very rich and newer interesting topological features when compared with the static topological case\cite{Kitagawa10,Kitagawa12,FloTop13}. For instance, periodically driven systems can give rise to Majorana-like edge states\cite{maj11}, chiral and counter-propagating edge states\cite{Usaj14}, among many others\cite{Fregoso13,Benito14,Thomas14,KHM16,Tetsuyuki16}. The emergence of edge states is protected by a conservation law or symmetry of the bulk system, this is the so-called bulk-edge correspondence\cite{Volovik2011}. The role played by the symmetries is fundamental to correctly describe the topological properties of these kind of systems, since they shed light about the topological invariant  that can be used to describe them\cite{Schnyder08,Graf2013}. Although great progress has been made in the topological classification of periodically driven gapful systems\cite{CAR2015,Nathan15}, the topological classification of gapless systems is yet incomplete. For instance, the topological properties of Dirac semimetals cannot be described by the topological invariants used for gapful systems\cite{Volovik2013}.

This is precisely the case of the topological modes in graphene. This material is a truly two dimensional crystal with extraordinary mechanical properties(that have given rise to very interesting phenomena in mechanically deformed graphene\cite{Olivaany14,Nosotros14,Nosotros214,tunoliva15,roman15}), which is known to have a non-trivial topological behavior not only in the static case\cite{HATSUGAI09,ZakPhase11,Nosotros214,Majgraphene15,Feilhauer15} but also in the periodically driven case\cite{Delplace13,Volkov16,Yan16,Roman17}. Among these interesting phenomena for the time-dependent case, we can cite chiral edge states\cite{Usaj14}, flat bands\cite{Roman17}, Majorana-like edge modes\cite{clement14}, and so on. 

Motivated by the previous discussion, we decided to study the emergence of edge states in periodically driven uniaxial strained armchair graphene nanoribbons (AGNs) using a tight binding approach within the Floquet theory. In particular and thinking on the possibility of achieving this system experimentally, we consider a spatial-periodic strain field. It is important to remark that in a previous work we considered the case of a periodically driven strained zigzag graphene nanoribbon\cite{Roman17} (ZGN). The case studied here is fundamentally different from the zigzag one. The first and maybe most important difference is that the AGN case can support gapless and gapful phases while the ZGN case can only support a gapless phase. This is a consequence of the static properties of strained graphene nanoribbons (see, for example, \cite{Pereira09,Nosotros214,Review17}). Second, the edge states nature of periodically driven strained AGN is completely different from the one observed in the ZGN case, see\cite{Roman17}. For instance, as we will see later on, for periodically driven AGN in the gapless phase, cosine-like edge states emerge, these states merge and coexist with the bulk bands. On the other hand, for the gapful phase, besides the edge sates that appear in the gap centered at zero quasienergy, another edge states emerge in other gaps. This is interesting since it has been proven that for periodically driven chiral system, fully gaps around zero and $\pm\pi$ quasienergies are topologically trivial from the Chern number point of view \cite{Michel16}. The implications of this fact are discussed below. 

To finish, the paper is organized as follows. In section \ref{model} we describe the theoretical model and the notation to be used while in section \ref{numerical} we analyze the quasienergy spectrum and the edge states for both the gapped and the gapless phases of our model. The topological properties of the edge states for both the gapless and the full gapped phase are discussed in section \ref{topological} via the symmetries of the time evolution operator. In section \ref{analytical} we study the topological nature of the edge states that emerge in the gapless phase via an effective Hamiltonian approach. Some conclusions are proposed in section \ref{conclusion}. Appendices \ref{A} and \ref{akxcero} include some calculations concerning the main text.

\section{Periodically driven strained graphene}
\label{model}
\begin{figure}
\includegraphics[scale=0.41]{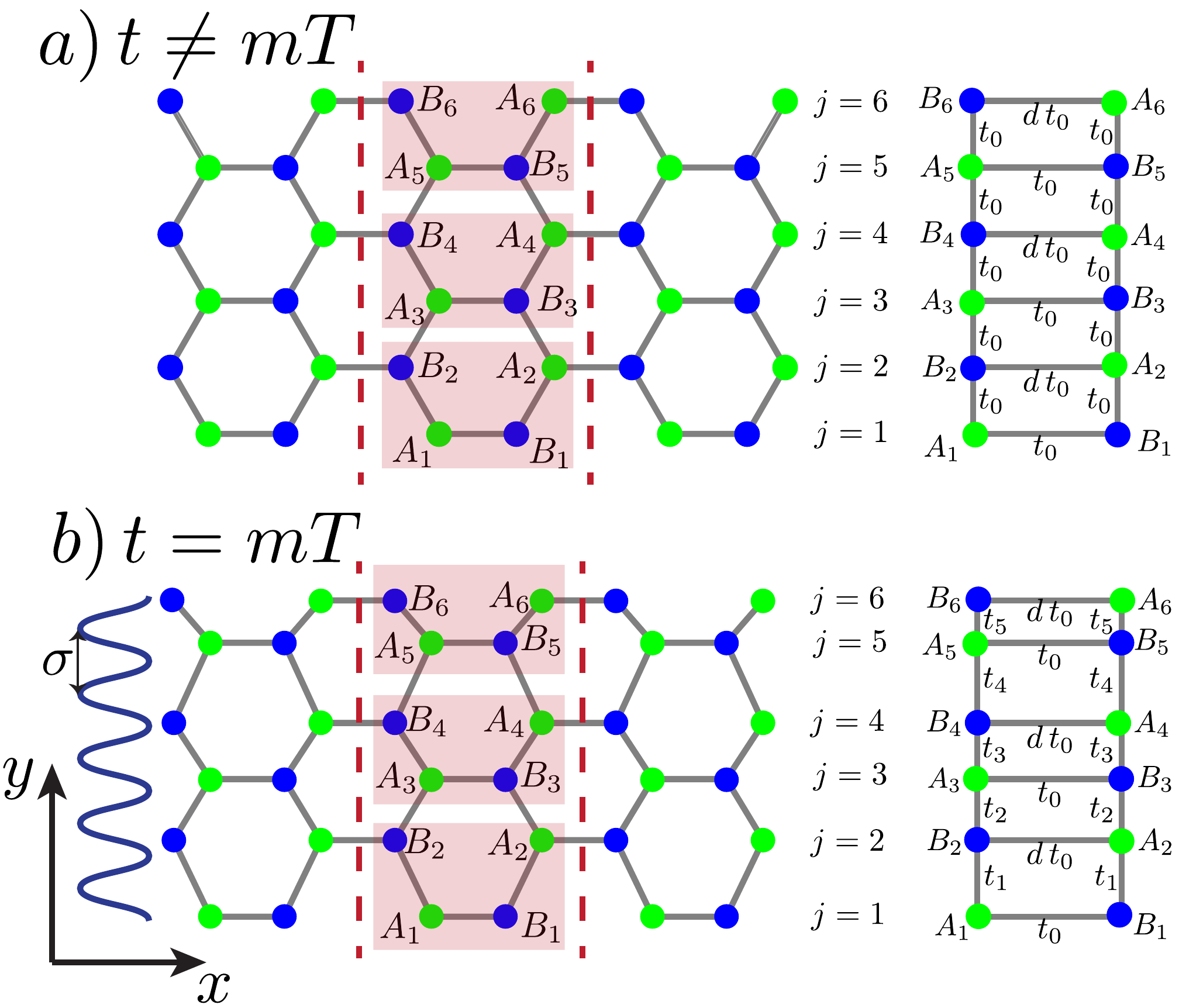}
\caption{(Color online). Schematic representation of the periodically delta driving layout for an armchair graphene nanoribbon.  For $t\neq T$, the strain field is turned off as shown in panel a). The strain field is turned on for $t=mT$ as can be seen in panel b). Here $T$ is the period of the delta driving. For uniaxial strain along the $y$-direction (in particular we consider a sinusoidal strain with wavelength $\sigma$), the strained armchair graphene nanoribbon can be mapped onto a one-dimensional effective system, which is represented by linear ladders on the left of the figure. Solid dots represent the position of carbon atoms. The shaded areas correspond to some unitary cells of pristine graphene, while the red dotted lines correspond to the limits of the unit cell in the $x$ direction. $j$ labels the rows in graphene, while A and B are used to denote each of the bipartite lattices\cite{Review17}. Carbons on each sublattice are indicated by green and blue.}
\label{layout}
\end{figure}
Consider a pristine armchair graphene nanoribbon (AGN) as the one shown in Fig. \ref{layout} b). Suppose that now we apply a spatially periodic strain field along the $y$-direction, given by
\begin{equation}
u(y_j)=\lambda\cos{\left(2\pi\sigma\,y_j+\phi\right)},
\label{strain}
\end{equation}
here $\lambda$ is the amplitude, $\sigma$ is the wavelength, and $\phi$ the phase of the strain field. $y_j$ are the positions of the carbon atoms along the $y$-direction inside the unit cell (see Fig. \ref{layout}, therein the unit cell of the system is indicated by dotted red lines).
The main advantage of uniaxial strain is the symmetry along the $x$-axis that allows to have a good quantum number associated, the quasimomentum along $x$-direction, that we denote by $k_x$. As a result, it is possible to decouple the two-dimensional system into an effective one-dimensional system\cite{Roche08,Nosotros14,Nosotros214}. In the tight binding limit, the electronic properties of graphene under an uniaxial strain field, as the one in Eq. (\ref{strain}), are described by the following one-dimensional (1D) effective Hamiltonian \cite{Nosotros214}
\begin{equation}
\begin{split}
H(k_x)=&\sum^{N/2-1}_{j=1} \gamma_0\left[d(k_x)a_{2j}^{\dag}b_{2j}+a_{2j-1}^{\dag}b_{2j-1}\right]\\
+&\sum_{j=1}^{N/2-1}\gamma_ja_j^{\dag}b_{j+1}+\mathrm{h.c.},
\label{HamStrain}
\end{split}
\end{equation}
where $d(k_x)=\exp{(3i\,k_x a_{c})}$, $k_x$ is the crystal momentum in the $x$-direction and $a_{c}$ is the interatomic distance between carbon atoms in unstrained graphene. In what follows, we will set $a_{c}=1$ and thus all quasimomenta $k_x$ are measured in units of $a_{c}$. Also, $N$ is the number of atoms per unit cell and $a_j$ ($b_j$) annihilates an electron at the site $j$ in the graphene's bipartite sublattice A (B), see Fig. \ref{layout}. The hopping parameters are given by,
\begin{equation}
\frac{\gamma_j}{\gamma_0}=\exp{\left[\beta \left(\sqrt{1-\lambda\sqrt{3}f(j,\sigma,\phi)+\lambda^2f^2(j,\sigma,\phi)}-1\right)\right]},
\label{hopping}
\end{equation}
where $\beta\approx3.37$ is the rate of decay \cite{Maurice} (Grun\"eissen parameter). The parameter $\gamma_0=2.7\,\text{eV}$ is the interatomic hopping parameter for pristine graphene that we will set as $\gamma_0=1$ in what follows, this is, all energies will be measured in units of $\gamma_0$. Finally the function $f(j,\sigma,\phi)$ is defined as, 
\begin{equation}
f(j,\sigma,\phi)=2\sin{\left(\sqrt{3}\pi\sigma/2\right)}\sin{\left[\sqrt{3}\pi\sigma(j+1/2)+\phi\right]}.
\label{function}
\end{equation}
The main features of this Hamiltonian have been described in a previous work, in the small strain's amplitude limit \cite{Nosotros214}. Let us make some remarks about the difference between considering zigzag or armchair graphene nanoribbons. As has been proven before, it is much easy to open a gap applying a strain field along the zigzag direction on an armchair graphene nanoribbon \cite{Review17}. Therefore, one expect that gaps emerge for certain parameters' values. In addition, one also expect edge states to be very different from the zigzag case\cite{Maksimov13}. Indeed this is the case as will be seen later on.
 
Once the Hamiltonian of an uniaxial strained AGN have been presented, we now introduce the time dependence to the model. In particular, we consider the case of a short-time strain impulse that can be approximated as a delta kicking. A graphic description of the driving layout is shown in Fig. \ref{layout}. Therein, we see that the deformation field is turned on at times $t=mT$ where $T$ is the driving period and $m$ is an integer number. The strain is turned off whenever that $t\neq mT$. That is, we consider a time-dependent Hamiltonian of the following form
\begin{equation}
H(k_x,t)=H_0(k_x)+\sum_{m}\left[H_1(k_x)-H_0(k_x)\right]\delta(t/T-m),
\label{tham}
\end{equation}
with the Hamiltonians $H_0$ and $H_1$ given by,
\begin{equation}
\begin{split}
H_0(k_x)=&\sum^{N/2-1}_{j=1} \gamma_0\left[d(k_x)a_{2j}^{\dag}b_{2j}+a_{2j-1}^{\dag}b_{2j-1}\right]\\
+&\sum_{j=1}^{N/2-1}\gamma_0a_j^{\dag}b_{j+1}+\mathrm{h.c.},
\label{H0}
\end{split}
\end{equation}
and
\begin{equation}
\begin{split}
H_1(k_x)=&\sum^{N/2-1}_{j=1} \gamma_0\left[d(k_x)a_{2j}^{\dag}b_{2j}+a_{2j-1}^{\dag}b_{2j-1}\right]\\
+&\sum_{j=1}^{N/2-1}\gamma_ja_j^{\dag}b_{j+1}+\mathrm{h.c.}
\label{H1}
\end{split}
\end{equation}
Even though the experimental realization of the proposed driving layout is experimentally challenging, there are some proposed experiments for similar situations\cite{Mishra2015,Agarwala16}. 

To study the quasienergy spectrum we construct the one-period time evolution operator of the system, defined as \cite{Rudner13},
\begin{equation}
U(k_x,T)\ket{\psi_{k_x}(t)}=\ket{\psi_{k_x}(t+T)},
\end{equation}
where $\ket{\psi_{k_x}(t)}$ is the time-dependent wave function of the system for a given quasimomentum $k_x$ along the $x$-axis. For the considered delta-kicking driving layout, $U(k_x,T)$ can be written in a very simple manner, namely,

\begin{equation}
\begin{split}
U(k_x,\tau)&=\mathcal{T}\exp{\left[-i\int_{0}^TH(k_x,t)\,dt/\hbar\right]}\\
 &=\exp{\left[-i\tau (H_1(k_x)-H_0(k_x))\right]}\exp{\left[-i\tau H_0(k_x)\right]},
\end{split}
\label{uop}
\end{equation}
where  $\mathcal{T}$ denotes the time ordering operator and $\tau=T/\hbar$. 

Even though the Hamiltonians $H_1(k_x)$ and $H_0(k_x)$ do not commute, we can study the eigenvalue spectrum of $U(k_x,\tau)$  via an effective Hamiltonian defined as,
\begin{equation}
U(k_x,\tau)=\exp{(-i\tau H_{\text{eff}}(k_x,\tau))} 
\end{equation}
The previous time evolution operator has eigenvalues $\exp{(-i\tau\omega)}$, where $\tau\omega$ are called the quasienergies of the system. They are defined up to integer multiples of $2\pi$. After introducing the time-dependence to the model, there are four free parameters: three owing to the strain field ($\lambda$, $\sigma$, and $\phi$) and one to the driving ($\tau$). 

One can study the system for a  wide range of parameters. Maybe the most important one is $\sigma$ since it controls the wavelength of the strain field. If this wavelength is incommensurate with respect to the graphene cell, the system is quasiperiodic resulting in a complex spectrum for the static undriven case \cite{Nosotros14,Review17}. However, since topological states are observed only when translational invariance holds \cite{RaoWeyl16}, here we study only commensurate cases.  In particular, we chose two different values for $\sigma$, namely, 1) $\sigma=1/\sqrt{3}$ and 2) $\sigma=0.5/\sqrt{3}$, setting $\phi=\pi\sigma$ for each case. We have chosen such $\sigma$ values since $\sigma=1/\sqrt{3}$ gives the smallest spatial period along the $y$-axis and the system is on a gapless phase around zero and $\pm\pi$ quasienergy in the bulk spectrum. While for $\sigma=0.5/\sqrt{3}$ we obtain the next size of the spatial period and the system is gapped around zero quasienergy in the bulk spectrum. 

For $\sigma=1/\sqrt{3}$ the supercell contains two rows of graphene in the $y$ direction, or in other words, four inequivalent carbon atoms in the supercell, since the hopping parameters just take two different vales, which are given by substituting the following expression
\begin{equation}
f(j,1/\sqrt{3},\phi)=2\lambda(-1)^{j}\cos{\left(\phi\right)},
\end{equation}
in Eq. (\ref{hopping}). On the other hand, for $\sigma=1/2\sqrt{3}$ the hopping parameters takes four different values, meaning that now the supercell has eight inequivalent atoms. Once again, the hopping parameters are given by substituting the following expression,
\begin{equation}
f\left(j,\frac{1}{2\sqrt{3}},\phi\right)=2 \lambda  \sin{\left[\frac{\pi}{2}(j+1/2)\right]}\sin{\left[\frac{\pi}{4}+(-1)^j \phi\right]},
\end{equation}
in Eq. (\ref{hopping}). 

\section{Quasienergy spectrum}
\label{numerical}

%
\begin{figure}
\includegraphics[scale=0.4]{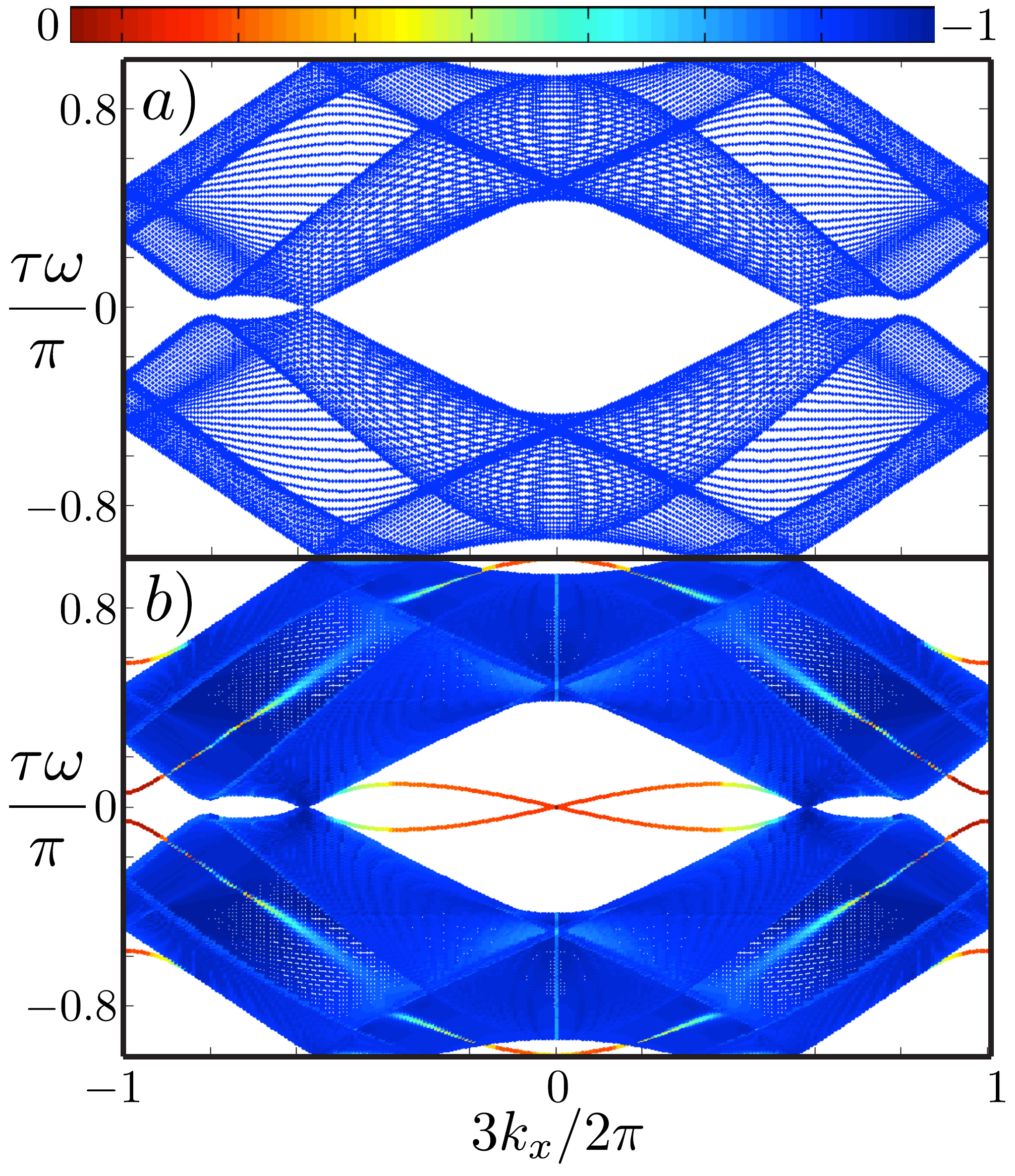}
\caption{(Color online). Gapless quasienergy band structure obtained from the numerical diagonalization of the matrix representation given by Eq. (\ref{uop}). The parameters used are $\sigma=1/\sqrt{3}$, $\lambda=0.2$, $\tau=\pi$, $\phi=\pi\sigma$, and $N=324$, for a) cyclic boundary conditions and b) fixed boundary conditions. In Panel a), note that no edge states appear since cyclic boundary conditions were used. In Panel b), two kinds of edge states emerge. Ones appear around zero quasienergy. The others are cosine-like edge states that merge and coexist with the bulk bands. The colors represent the inverse participation ratio. For red color the states are highly localized, whereas for blue color they are totally delocalized.}
\label{bs1}
\end{figure}
\begin{figure}
\includegraphics[scale=0.41]{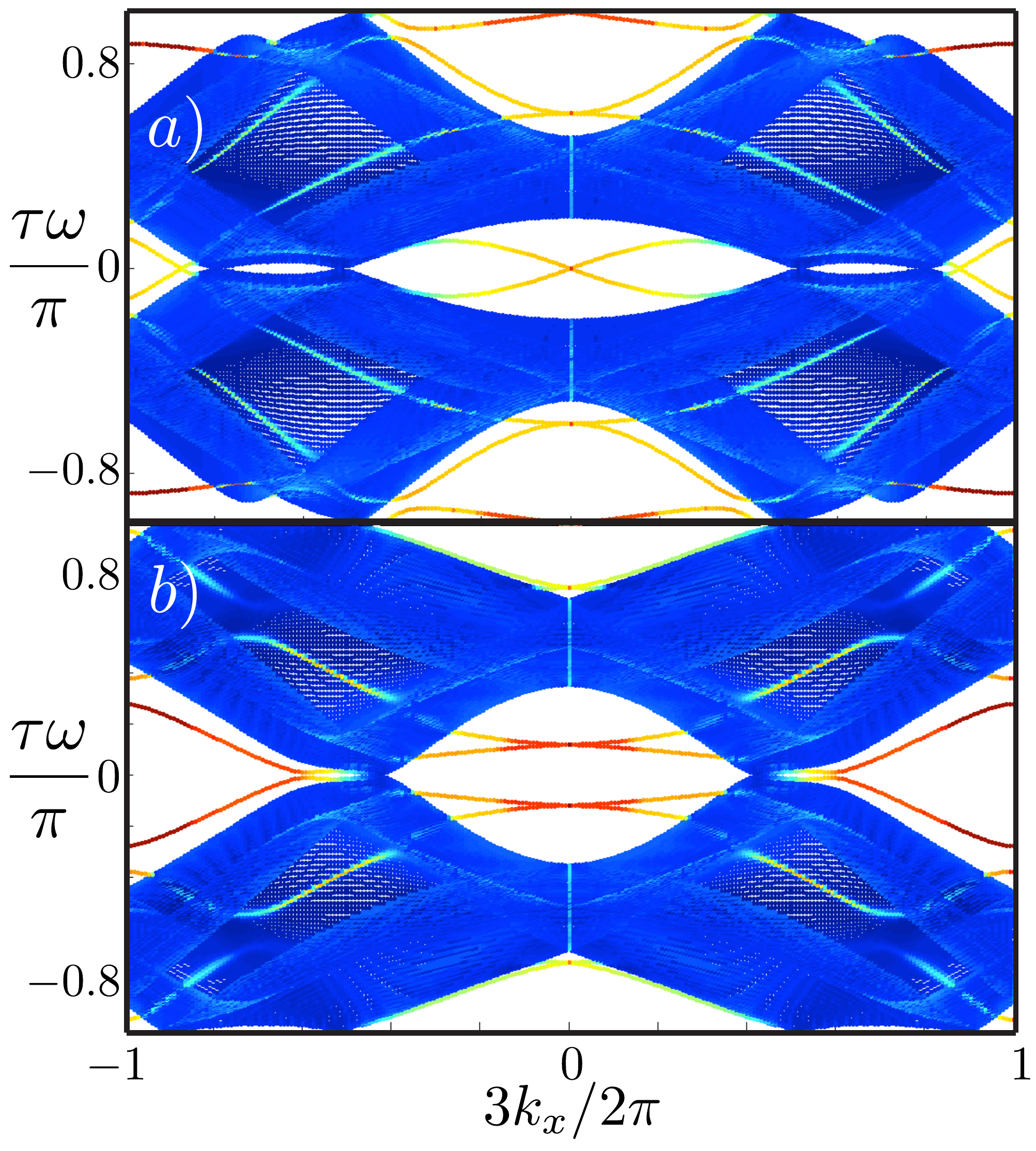}
\caption{(Color online). Gapless quasienergy band structure obtained from the numerical diagonalization of the matrix representation of Eq. (\ref{uop}). The parameters used are $\sigma=1/\sqrt{3}$, $\lambda=0.2$, $\tau=\pi$, $N=324$, and using fixed boundary conditions for a) $\phi=0$ and b) $\phi=\pi\sigma/2$. Note that for a) and b) panels the edge states deeply penetrate into the bulk bands. Also, for panel b), edge states around zero quasienergy are decoupled. It was used the same color code as in Fig. \ref{bs1} to display localization of each mode.}
\label{bs3}
\end{figure}
\begin{figure}
\includegraphics[scale=0.422]{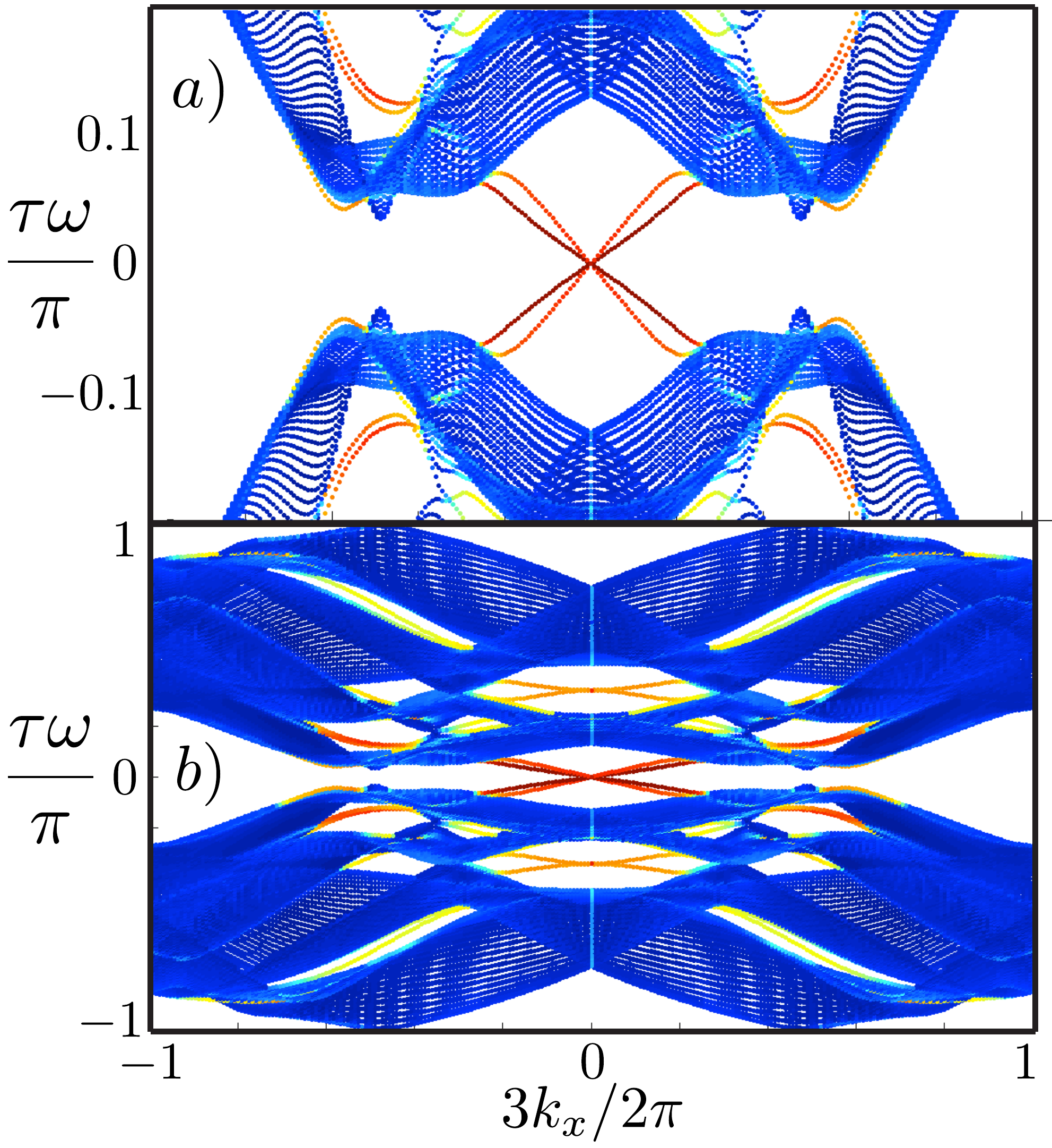}
\caption{(Color online). Gapful quasienergy band structure obtained from the numerical diagonalization of the matrix representation of Eq. (\ref{uop}) for $\sigma=1/2\sqrt{3}$, $\lambda=0.2$, $\tau=\pi$, $\phi=\pi\sigma$, and $N=324$ using fixed boundary conditions. In panel a) we show an amplification around zero quasienergy to highlight the zero quasienergy edge modes. Observe that edge states also emerge at other gaps. The same color code as in Fig. \ref{bs1} was used to display localization of each mode.}
\label{bs2}
\end{figure}

In this section we will study the quasienergy spectrum and the emergence of topological edge states. 
We start by building the matrix representation of $U(k_x,\tau)$ as described in Eq. (\ref{uop}). Then, their eigenvalues as a function of $k_x$ are obtained by numerical calculations for a finite system. For all numerical cases presented here we will consider a system with $N=324$ . It is worthwhile to note that for $\sigma=1/\sqrt{3}$ or $\sigma=1/2\sqrt{3}$ the system becomes periodic in the $y$-direction. Therefore, by applying cyclic boundary conditions along the $x$- and $y$-directions, the quasienergy spectrum can be also obtained by Fourier transforming the Hamiltonians $H_0(k_x)$ and $H_1(k_x)$ and then substituting them into Eq. (\ref{uop}). No edge states appear in the quasienergy dispersion relation obtained by using this method but allows to compare numerical with analytical results. To observe edge states, here we needed to perform calculations in real space for the $y$-direction. 

Let us start by studying the quasienergy spectrum for $\sigma=1/\sqrt{3}$, in other words, the gapless quasienergy spectrum. In Fig. \ref{bs1} we show the quasienergy band structure for $\lambda=0.2$, $\tau=\pi$, $\phi=\pi\sigma$, and $\sigma=1/\sqrt{3}$. In panel a) we have used cyclic boundary conditions , whereas for panel b) the boundary conditions were changed to fixed. Observe that the main difference between panel a) and b) is that in b) edge states emerge. The colors in the figure represent the logarithm of the normalized inverse participation ratio (IPR) as, defined in Ref. \cite{Nosotros14},
\begin{equation}
 \alpha(E)=\frac{\ln \sum_{j=1}^{N} |\psi(j)|^{4}}{\ln N}
\end{equation}
where $\psi(j)$ is the eigenfunction at site $j$ for an energy $E$. The IPR is a measure of the wave function localization. The closer to zero the IPR (red color in figures) the more localized the wave function is. Delocalized or extended wave functions are labeled by blue color in the figures, and correspond to $\alpha(E)$ tending to $-1$. 

It is interesting to note that even though the system is on a gapless phase, edge states appear. Moreover, two kinds of edge states are observed in Fig. \ref{bs1}. Ones around zero quasienergy, which are degenerate at $k_x=0$, but decouple and delocalize as they move away from that point, following a cosine-like dispersion. The other ones also have a cosine-like dispersion and are degenerated at $k_x=0$ at $\pm\pi$ quasienergy. As $k_x$ moves away from that point, edge states decouple and, eventually merge with the bulk bands without being totally delocalized states. We have numerically checked that they are localized near opposite edges of the AGN. Also, we stress out that the quasienergy spectrum strongly depends upon the phase $\phi$ of the strain field. To see this point, in Fig. \ref{bs3} we present the quasienergy band structure for: $\phi=0$ (panel a) and $\phi=\pi\sigma/2$ (panel b). In panel a), it can be seen that edge states are quite similar to the ones shown in Fig. \ref{bs1} b). However, the edge states in Fig. \ref{bs3} a) deeply penetrate into the bulk bands. Whereas, the edge states in Fig. \ref{bs3} b) do not touch neither at zero nor at $\pm\pi$ quasienergies when $k_x=0$, instead a finite gap between them has been opened. Besides, they decouple into four bands around zero quasienergy, and also deeply penetrates into the bulk states. The strong dependence of the quasienergy band structure on $\phi$ can be explained as follows. Basically, the phase $\phi$ determines how the strain pattern matches with the edges of the AGN, a fact that has been proven to be crucial in the topological properties of similar systems\cite{SSH79,SSHdriven15}. In the next section, we topologically characterize the edge states that emerge for $\phi=\pi\sigma$, showing that they are at least weak topologically non-trivial. 

Since topological edge states are very robust to small perturbations, we expect edge states not to be destroyed by a full gap in the bulk spectrum. To confirm this assertion, in Fig. \ref{bs2} we plot the quasienergy band structure in a gapped phase of the system, this is, we used the same conditions as in Fig. \ref{bs2} but using $\sigma=1/2\sqrt{3}$ for fixed boundary conditions. As can be seen the quasienergy spectrum is on a gapped phase. Note that four edge states emerge around zero quasienergy. They merge in a single point at $k_x=0$, as happens in Fig. \ref{bs1}. Besides that, other edge states emerge always that a partly gap appears on the quasienergy spectrum. As will be shown in the next section, our model possesses chiral symmetry and thus edges states that appear in the gap around zero quasienergy are topologically trivial  from the point of view of the Chern number\cite{Michel16}, although they can have a weak topological nature \cite{Yoshimura14,Ho14,WeakTop15}. However, edge states at other gaps (this is, full gaps not centered around zero or $\pm\pi$ quasienergy) can be topologically non-trivial\cite{Michel16}. 

\section{Topological properties of edge states}
\label{topological}

Before entering the study of the topological properties of the system, let us write the Fourier transformed version of the Hamiltonians $H_1(k_x)$ and $H_0(k_x)$ when periodic boundary conditions are used in the $y$ direction. Then it is possible to fully write the Hamiltonians $H_1(k_x)$ and $H_0(k_x)$ in reciprocal space by taken into account the new periodicity in $y$-direction. This leads to a new quantum number $k_y$, from where it follows that $H_1(k_x)$ and $H_0(k_x)$ can be simplified using a suitable Fourier transform. In fact, it can be proven that such Hamiltonians are reduced to a $4Q\times4Q$ matrix dependent on ${\bf k}=(k_x,k_y)$, where $4Q$ is the number of inequivalent rows in the $y$ direction. Notice that $Q$ is related to $\sigma$ as $\sigma=P/(\sqrt{3}Q)$, since in Eq. (\ref{strain}) the positions $y_j$ are evaluated at graphene's sites along a zigzag path, where atoms are separated by distances $\sqrt(3)/2$. For the case $\sigma=1/\sqrt{3}$, the matrices have its lowest size since $Q=1$, therefore Hamiltonians $H_1(\mathbf{k})$ and $H_0(\mathbf{k})$ are $4\times4$ matrices. This is the most simple case and can be studied analytically as will be done in the next section.

Yet one can make further progress by writing the Hamiltonian for a general $Q$ in the chiral basis, i.e. in a basis such that all sites in the $A$ sublattice appear as the first entries in the vector, then followed by its corresponding counterparts in the $B$ sublattice \cite{Review17}. Then one obtains,
\begin{equation}
\mathbb{H}_l(\mathbf{k})=\left[
\begin{array}{cc}
0&\tilde{\mathbb{H}}_l(\mathbf{k})\\
\tilde{\mathbb{H}}^{\dag}_l(\mathbf{k})&0
\end{array}
\right],
\label{H1H0Bulk}
\end{equation}
where $l=0,1$ and the tilde indicates $2Q\times2Q$ matrices. The explicit form of $\tilde{\mathbb{H}}_l(\mathbf{k})$ is given in appendix \ref{A} for the particular case of  $\sigma=1/\sqrt{3}$. The perturbation $\delta \mathbb{H}(k_y)=\mathbb{H}_1(\mathbf{k})-\mathbb{H}_0(\mathbf{k})$ is simply written as,
\begin{equation}
\delta \mathbb{H}(k_y)=\left[
\begin{array}{cc}
0&\delta\tilde{\mathbb{H}}(k_y)\\
\delta\tilde{\mathbb{H}}^{\dag}(k_y)&0
\end{array}
\right],
\label{dHBulk}
\end{equation}
where $\delta\tilde{\mathbb{H}}(k_y)= \tilde{\mathbb{H}}_1(\mathbf{k})-\tilde{\mathbb{H}}_0(\mathbf{k}) $.  The explicit form of these matrices for  $\sigma=1/\sqrt{3}$ is given in appendix \ref{A}. Notice that $k_x$ cancels out as the perturbed and unperturbed Hamiltonians have the same symmetry in the $x$ axis.

For studying the topological properties of our model we start by looking at the symmetries of the Hamiltonians $\mathbb{H}_1(\mathbf{k})$ and $\mathbb{H}_0(\mathbf{k})$. Note that such Hamiltonians fulfill the following condition,
\begin{equation}\label{ChiralProperty}
\Gamma \mathbb{H}_l(\mathbf{k})\Gamma=-\mathbb{H}_l(\mathbf{k})
\end{equation}
where $l=0,1$ and $\Gamma$ is the so called chiral operator. $\Gamma$ is an unitary operator which can be represented, in the chiral basis, as
\begin{equation}
\Gamma=
\left[\begin{array}{cc}
\mathbb{I}_{2Q\times2Q}&0\\
0&-\mathbb{I}_{2Q\times2Q}
\end{array}\right]
\end{equation}
with the property that $\Gamma^2=\mathbb{I}_{4Q\times4Q}$. 
As a consequence of Eq.(\ref{ChiralProperty}), the Fourier transformed version of the time-dependent Hamiltonian Eq. (\ref{tham}) possesses the following property
\begin{equation}
\Gamma \mathbb{H}(\mathbf{k},t)\Gamma=-\mathbb{H}(\mathbf{k},-t)
\label{gammat}
\end{equation}

From Eq. (\ref{gammat}) it follows that the Hamiltonian Eq. (\ref{tham}) is chiral, therefore, the time evolution operator Eq. (\ref{uop}) must satisfy the following condition
\begin{equation}
\Gamma \mathbb{U}(\tau)\Gamma=\mathbb{U}^{-1}(\tau)=\mathbb{U}^{\dag}(\tau)=\mathbb{U}(-\tau),
\label{ugamma}
\end{equation}
where the time evolution operator is now given by,
\begin{equation}
\mathbb{U}(\tau)=e^{-i\tau\delta\mathbb{H}(\mathbf{k})}e^{-i\tau\mathbb{H}_0(\mathbf{k})}.
\label{Uper}
\end{equation}
Observe that by using Eq. (\ref{ugamma}), $\mathbb{U}(-\tau)$ is now written as, 
\begin{equation}
\mathbb{U}(-\tau)=e^{i\tau\mathbb{H}_0(\mathbf{k})}e^{i\tau\delta\mathbb{H}(\mathbf{k})}.
\end{equation}
which is the same result that one obtains directly from the time ordering operator that appears in Eq. (\ref{uop}). 

Due to the condition Eq. (\ref{ugamma}), the quasienergy spectrum is symmetric respect to reflections along the $k_x$-axis, as confirmed in Figs. \ref{bs1}, \ref{bs3}, and \ref{bs2}. Moreover, the chiral symmetry for fully gapped systems in two dimensions imposes the vanishing of the topological invariant at full gaps centered around zero and $\pm\pi$ quasienergy\cite{Michel16}. Other gaps can be topologically non-trivial\cite{Michel16}. As was mentioned before, for the case $\sigma=1/2\sqrt{3}$ the topological invariant is zero but edge states emerge in the gap centered at zero quasienergy, see Fig. \ref{bs2}. This implies that a different topological invariant is needed to topologically characterize the system or that the edge states are topologically weak \cite{Yoshimura14}. On the other hand, for the gapless phase of the system, it is not clear the topology nature of the edge states \cite{Michel16}. Therefore, to topologically characterize the edge states that emerge in the gapless phase of our system (see Fig. \ref{bs1}), we will study the topological properties of a one-dimensional slice of the two-dimensional system, {\it i.e}, we consider the case $k_x=0$. As we will see in what follows, this one dimensional slide is topologically non-trivial.

\section{Analytical study of the topology for $\sigma=1/\sqrt{3}$ at $k_x=0$}
\label{analytical}

\begin{figure}
\includegraphics[scale=0.342]{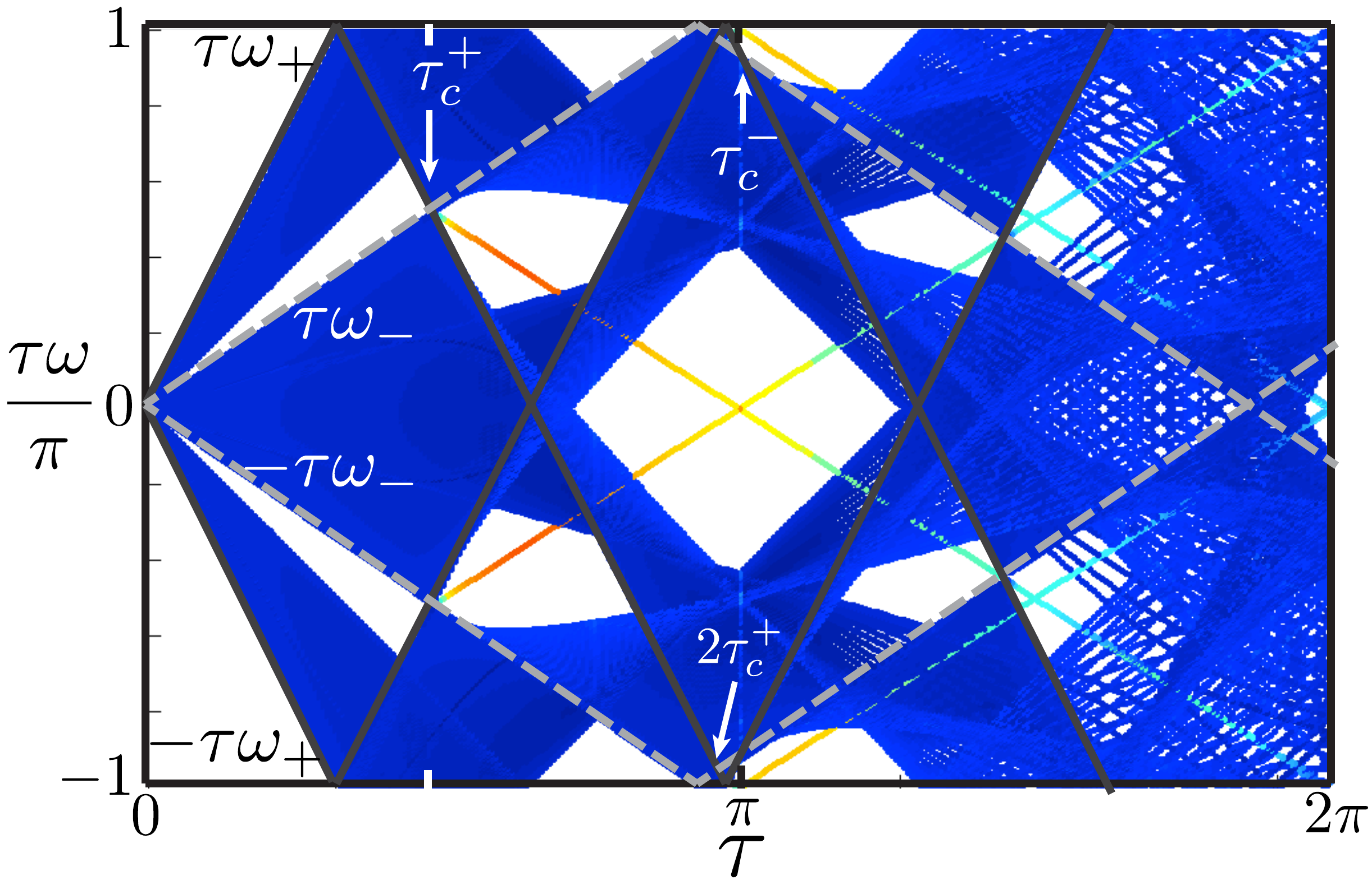}
\caption{(Color online). Quasienergy band spectrum for $k_x=0$ as a function of $\tau$ obtained by numerical diagonalization of Eq. (\ref{uop}) using $N=324$, $\sigma=1/\sqrt{3}$, $\lambda=0.2$, $\phi=\pi\sigma$, and fixed boundary conditions compared with the band edges obtained from the analytical calculation given by Eq. (\ref{omp}) and Eq. (\ref{tauc}). The quasienergy band edge $\tau\omega^{+}$ ($\tau\omega^{-}$) is denoted by solid black (dotted gray) lines. Note that such band edges cross each other for the first time at $\tau=\tau_{c}^{\pm}$ at quasienergies different from zero or $\pm\pi$. It is interesting that for $\tau=\tau_c^{+}$ edge states emerge, which is not the case for $\tau_c^{-}$. Also, observe that the edge states are no flat bands but they depend linearly on $\tau$. See section \ref{analytical}.}
\label{kxcer}
\end{figure}

We begin studying the emergence of edge states for the gapless case, obtained for $\sigma=1/\sqrt{3}$. For doing that,  as seen in Fig. \ref{kxcer}, note that edge states will emerge for the first time when the lower and upper quasienergy band edges cross each other\cite{Winkler17} as $\tau$ is increased from zero (keeping the other parameters fixed). Band edges correspond to extremal values of the quasienergy spectrum for the time evolution operator given by Eq. (\ref{Uper}). It is easy to see that such extremal values are reached when the Hamiltonians $\mathbb{H}_0$ and $\delta\mathbb{H}$ commute. Let us denote by $k_x^{*}$ and $k_y^{*}$ the points where this happens. From Eq. (\ref{commutator}), we can readily obtain that $k_x^{*}=3n\pi/2$ and $k_y^{*}=n\pi/\sqrt{3}$, where $n$ is an integer number. Since we are interested in a one-dimensional slice of our system, we  first consider $k_x=0$ and then study the quasienergy spectrum as a function of $k_y$. As is proven in appendix \ref{akxcero}, for $k_x=0$, the time evolution operator becomes block diagonal,
\begin{equation}\label{uprime}
\mathbb{U^{\prime}}(\tau)=
\left[\begin{array}{cc}
e^{i\tau \delta\tilde{h}(k_y)}e^{i\tau \tilde{h}_0(k_y)}&0\\
0&e^{-i\tau \delta\tilde{h}(k_y)}e^{-i\tau \tilde{h}_0(k_y)}
\end{array}\right]
\end{equation}
where $\mathbb{U}^{\prime}(\tau)$ is the time evolution operator Eq. (\ref{Uper}) written in the basis where $\mathbb{H}_0$ is diagonal, see the appendix for details. $\delta\tilde{h}(k_y)$ and $\tilde{h}_0(k_y)$ can be written as follows,
\begin{equation}
\begin{split}
\delta\tilde{h}(k_y)&=\delta h(k_y)\,\delta\hat{\mathbf{h}}\cdot\mathbf{\sigma},\\
\tilde{h}_0(k_y)&=\mathbb{I}_{2\times2}+2\cos{\left(\sqrt{3}k_y/2\right)}\sigma_z,
\end{split}
\end{equation}
where $\mathbf{\sigma}=(\sigma_x,\sigma_y,\sigma_z)$, $\sigma_i$ ($i=x,y,z$) is the $2\times2$ Pauli matrix, $\delta\hat{\mathbf{h}}=\delta\mathbf{h}/\delta h$, and the components of $\delta\mathbf{h}$ are given by,
\begin{equation}
\begin{split}
\delta h^{(y)}&=(\gamma_1-\gamma_2)\sin{\left(\sqrt{3}k_y/2\right)},\\
\delta h^{(z)}&=(\gamma_1+\gamma_2-2)\cos{\left(\sqrt{3}k_y/2\right)},
\end{split}
\end{equation}
also we define the norm $\delta h(k_y)$ as 
\begin{equation}
\delta h(k_y)=\sqrt{\left[\delta h^{(y)}\right]^2+\left[\delta h^{(z)}\right]^2}.
\end{equation}
Now, it is possible to analytically obtain the quesienergies of $\mathbb{U^{\prime}}(\tau)$ by studying only one of the diagonal blocks in Eq. (\ref{uprime}). We will do that via an effective Hamiltonian defined as,
\begin{equation}
e^{i\tau \tilde{h}_{\text{eff}}(k_y)}=e^{i\tau \delta\tilde{h}(k_y)}e^{i\tau \tilde{h}_0(k_y)}
\end{equation}
For calculating $\tilde{h}_{\text{eff}}$ we use the addition rule of SU(2). After some algebraic operations detailed in the appendix, one gets,
\begin{equation}
\tilde{h}_{\text{eff}}(k_y)=\mathbb{I}_{2\times2}+\Omega(k_y)\hat{\mathbf{h}}_{\text{eff}}\cdot\mathbf{\sigma}
\label{heff}
\end{equation}
where $\hat{\mathbf{h}}_{\text{eff}}$ is a unit vector defined in appendix \ref{akxcero}, and $\tau\Omega(k_y)$ satisfies,
\begin{equation}
\begin{split}
&\cos{\left[\tau\Omega(k_y)\right]}=\cos{(\tau\delta h)}\cos{\left[2\tau\cos{\left(\sqrt{3}k_y/2\right)}\right]}\\
&-\hat{e}_z\cdot\delta\hat{\mathbf{h}}\sin{(\tau\delta h)}\sin{\left[2\tau\cos{\left(\sqrt{3}k_y/2\right)}\right]}.
\end{split}
\label{omegakx}
\end{equation}
The quasienergies are given by the eigenvalues of the time evolution operator, denoted by $\pm\tau\omega(k_y)$, from where,
\begin{equation}
\tau\omega(k_y)=\tau\left[\pm1+\Omega(k_y)\right].
\label{omegaeff}
\end{equation}

The topological information of the system for $\sigma=1/\sqrt{3}$ and $k_x=0$ is now contained in the effective Hamiltonian (\ref{heff}). It is illustrative to obtain the conditions for having edge states before studying the topological properties of the Hamiltonian (\ref{heff}). Since edge states will emerge when the lower and upper band edges cross each other for the first time, we begin by obtaining such band edges. As was mentioned before, the extreme values of the quasienergy spectrum are found at $k_x=k_x^*=0$ and $k_y=k_y^*=0$. After substituting such values in Eq. (\ref{omegakx}) one gets,
\begin{equation}
\tau\omega_{\pm}=\tau(\pm1+\gamma_1+\gamma_2).
\label{omp}
\end{equation}
Now, the condition for having quasienergy band edges crossings at the critical values $\tau=\tau_c$ are
given by,
\begin{equation}
\tau_{c}(\omega_{+}\pm\omega_{-})=2\pi.
\end{equation}
By using Eq. (\ref{omp}) we obtain,
\begin{equation}\label{tauc}
\begin{split}
\tau_{c}^{+}&=\frac{\pi}{\gamma_1+\gamma_2},\\
\tau_{c}^{-}&=\pi.
\end{split}
\end{equation}
All other band crossings are given by $m\tau_c^{\pm}$, where $m$ is an integer number. 
To shed light on the previous analysis, it is meaningful to compare it with numerical results. In Fig. \ref{kxcer}, we plot the quasienergy spectrum for $k_x=0$ as a function of $\tau$ obtained by numerical diagonalization of Eq. (\ref{uop}) using $N=324$, $\sigma=1/\sqrt{3}$, $\lambda=0.2$, $\phi=\pi\sigma$, and fixed boundary conditions. Therein, the band edges $\pm\tau\omega^{+}$ ($\pm\tau\omega^{-}$) calculated from Eq. (\ref{omp}) are denoted by solid black (dotted gray) lines. The critical values of $\tau$ obtained from Eq. (\ref{tauc}) are displayed as well. The agreement between the numerical and analytical calculations is excellent.

Notice in Fig. \ref{kxcer} how the edges $\tau\omega^{\pm}$ start at zero for $\tau=0$. As $\tau$ is increased, $\tau\omega^{+}$ and $\tau\omega^{-}$ grow linearly but with different slopes. Since $\omega^{+}\geq\omega^{-}$, $\omega^{+}$ reaches first the limit of the Floquet zone. Then, as $\tau$ is further increased, $\omega^{+}$ and $\omega^{-}$ cross each other at $\tau_{c}^{+}$. By further increasing $\tau$ a new band edge crossing emerges at $\tau_c^{-}$. Interestingly, the only crossings that produce edge states are the ones where $\tau=m\tau_{c}^{+}$. For the other crossings, (at $m\tau_c^{-}$) no gap is opened, hence no edge states emerge. Observe that for $\tau_c^{+}$ the band edge crossings do not occur at zero or $\pm\pi$ quasienergy, instead, they cross each other at,
\begin{equation}
\begin{split}
\tau_c^{+}\omega_{\pm}&=\pi\pm\frac{\pi}{\gamma_1+\gamma_2}.\\
\end{split}
\end{equation}
Unlike the case of driven uniaxial strained zigzag graphene nanoribbons\cite{Roman17}, the quasienergy at which these edge states emerge is different from zero or $\pm\pi$. Moreover, the edge states for the case  considered here are not flat bands but unidirectional edge states, meaning that the quasienergy of such states grows linearly with $\tau$, see Fig. \ref{kxcer}. Note that as $\tau$ is increased, edge states start to delocalize. When $\tau$ reaches $2\pi$ they are almost completely extended, as seen in Fig. \ref{kxcer} by looking at the colors that represent the normalized inverse participation ratio.  

\begin{figure}
\includegraphics[scale=0.4]{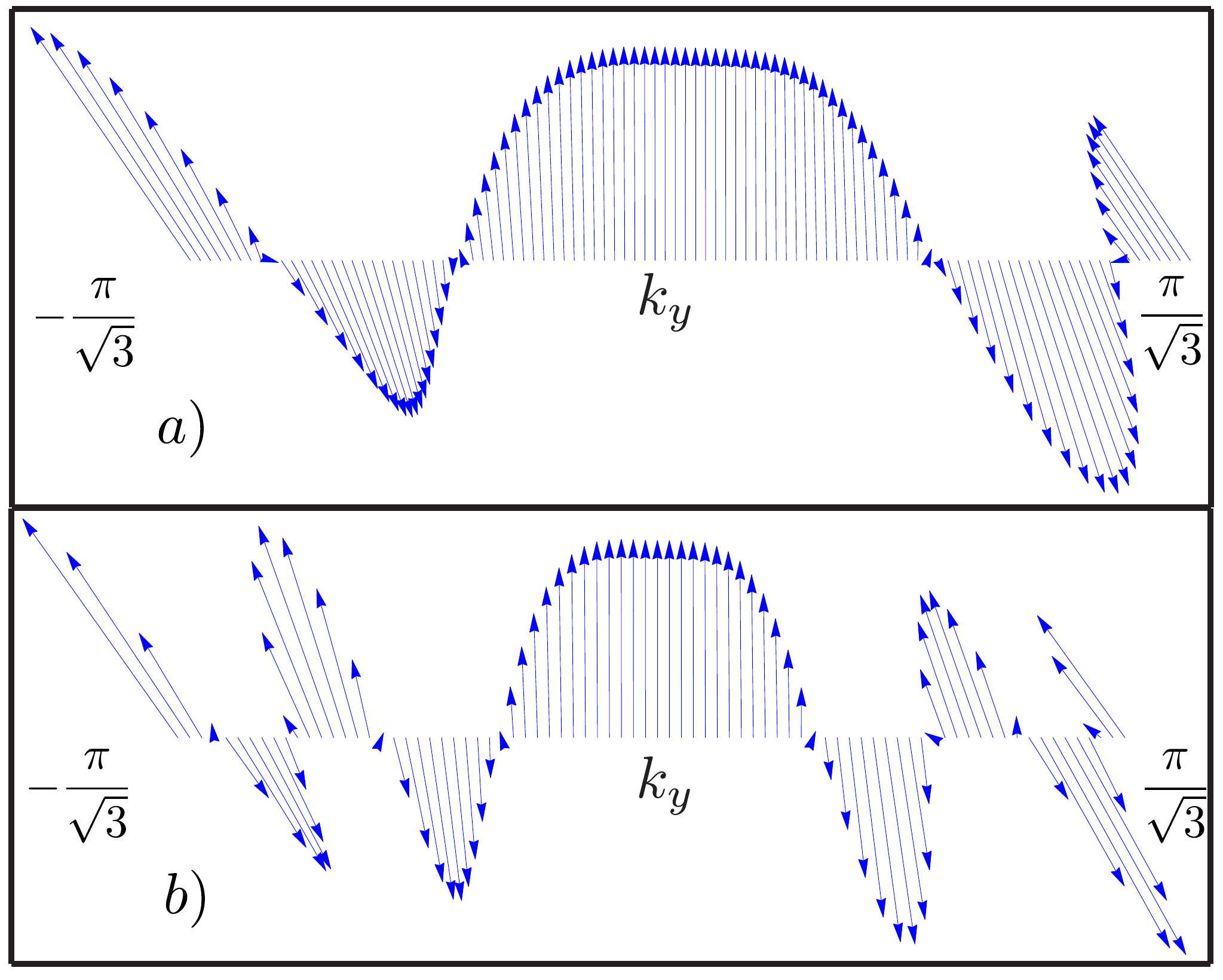}
\caption{(Color online). The winding of the vector $\hat{\mathbf{h}}_{\text{eff}}(k_y)$ for $k_x=0$, $\lambda=0.2$, $\sigma=1/\sqrt{3}$, and $\phi=\pi\sigma$ using $\tau=\pi$ for panel a) and $\tau=2\pi$ for panel b). The winding number for panel a) is $2$, whereas for panel b) is $4$. Observe that by increasing the driving period $\tau$, the winding number is also increased.}
\label{winding}
\end{figure}

To finish this section, we will show that these edge states for $\tau$ and $\lambda$ fixed are topologically non-trivial. This is done by studying the effective Hamiltonian Eq. (\ref{heff}). For having topologically non-trivial properties the winding number of the unitary vector $\hat{\mathbf{h}}_{\text{eff}}(k_y)$ around the origin must be non-vanishing. This is confirmed graphically in Fig. \ref{winding}, therein, we show the winding of $\hat{\mathbf{h}}_{\text{eff}}(k_y)$ for $k_x=0$, $\lambda=0.2$, $\phi=\pi\sigma$, and $\sigma=1/\sqrt{3}$ for a) $\tau=\pi$ and b) $\tau=2\pi$. As can be seen, the winding number for this particular case is 2 for panel a) and 4 for panel b). Hence, the system is topologically non-trivial for a one-dimensional slide at $k_x=0$. This means that the edge states observed in Fig. \ref{bs1} b) have a topologically weak nature.

\section{Conclusions}
\label{conclusion}

We have observed the emergence of topological edge states in periodically driven uniaxial strained AGN. The system has a gapped phase (for $\sigma=1/\sqrt{3}$) and a gapless phase (for $\sigma=1/2\sqrt{3}$). For the gapped phase, highly localized edge states were found around zero quasienergy, which, due to the chirality of the system, must be topologically trivial from the Chern number point of view\cite{Michel16}. However, this phase also exhibits edge states at gaps at quasienergies different from zero or $\pm\pi$ that could be topologically non-trivial, although a more detailed analysis is required. On the other hand, for the gapped phase of the system, we found the necessary conditions for the emergence of the edge states. Additionally, by studying a one-dimensional slide of the case system at $k_x=0$, we were able to analytically obtain the quasienergy spectrum of such slide, since for this case the time evolution operator can be effectively describe by a block diagonal $4\times4$ matrix, in other words, we obtained the effective Hamiltonian for $k_x=0$. After looking at the topological properties of the effective Hamiltonian, we found that for this slice the edge states are topologically non-trivial. Although, a deeper analysis is needed, we can say that the edge states observed in the case $\sigma=1/\sqrt{3}$ have, at least, a topologically weak nature. We hope that our work motives further research about the cases presented here. 


This project was supported by DGAPA-PAPIIT Project 102717. P. R.-T. acknowledges financial support from Consejo Nacional de Ciencia y Tecnolog\'ia (CONACYT) (M\'exico). We gratefully thank M. Fruchart for helpful discussions.

\appendix
\section{}
\label{A}

Let us now compute the unitary operator $\mathbb{U}(\tau)$. Before entering into the detailed calculation, we first define the Hamiltonians $\mathbb{H}_0$ and $\mathbb{H}_1$,
\begin{equation}
\mathbb{H}_l(\mathbf{k})=\left[
\begin{array}{cc}
0&\tilde{\mathbb{H}}_l(\mathbf{k})\\
\tilde{\mathbb{H}}^{\dag}_l(\mathbf{k})&0
\end{array}
\right],
\end{equation}
the label $l$ can take the values $l=0,1$, and 
\begin{equation}
\begin{split}
\tilde{\mathbb{H}}_0(\mathbf{k})&=\left[
\begin{array}{cc}
1&1+e^{-i\,\sqrt{3}k_y}\\
1+e^{i\,\sqrt{3}k_y}&e^{-i\,3k_x}
\end{array}
\right],\\
\tilde{\mathbb{H}}_1(\mathbf{k})&=\left[
\begin{array}{cc}
1&\gamma_1+\gamma_2e^{-i\,\sqrt{3}k_y}\\
\gamma_1+\gamma_2e^{i\,\sqrt{3}k_y}&e^{-i\,3k_x}
\end{array}
\right].
\end{split}
\end{equation}
Thus, the perturbing Hamiltonian to $H_0$ is defined as $\delta\mathbb{H}=\mathbb{H}_1(\mathbf{k})-\mathbb{H}_0(\mathbf{k})$, which takes the form,
\begin{equation}
\delta\mathbb{H}(k_y)=\left[
\begin{array}{cc}
0&\delta\tilde{\mathbb{H}}(k_y)\\
\delta\tilde{\mathbb{H}}(k_y)&0
\end{array}
\right],
\end{equation}
with
\begin{equation}
\delta\tilde{\mathbb{H}}(k_y)=\left[
\begin{array}{cc}
0&\delta_1+\delta_2e^{-i\,\sqrt{3}k_y}\\
\delta_1+\delta_2e^{i\,\sqrt{3}k_y}&0
\end{array}
\right],
\end{equation}
where $\delta_{1,2}=\gamma_{1,2}-1$. Note that the unperturbed and perturbed Hamiltonians do not commute. In fact, we have
\begin{equation}
\left[\delta\mathbb{H}(k_y),\mathbb{H}(\mathbf{k})\right]=\left[
\begin{array}{cc}
\tilde{\mathbb{C}}(\mathbf{k})&0\\
0&\tilde{\mathbb{C}}(\mathbf{k})
\end{array}
\right],
\label{commutator}
\end{equation}
where
\begin{equation}
\begin{split}
\tilde{\mathbb{C}}(\mathbf{k})&=2i(\gamma_1-\gamma_2)\sin{(\sqrt{3}k_y)}\sigma_z\\
&+2\left[\delta_1\sin{\left(\frac{3k_x}{2}\right)}+\delta_2\sin{\left(\frac{3k_x}{2}-\sqrt{3}k_y\right)}\right]\\
&\times\sin{\left(\frac{3k_x}{2}\right)}\sigma_x+\\
&2\left[\delta_1\cos{\left(\frac{3k_x}{2}\right)}+\delta_2\cos{\left(\frac{3k_x}{2}-\sqrt{3}k_y\right)}\right]\\
&\times\sin{\left(\frac{3k_x}{2}\right)}\sigma_y
\end{split}
\label{comm}
\end{equation}
where $\sigma_{x,y}$ are the $2\times2$ Pauli matrices and $\delta_{1,2}=\gamma_{1,2}-1$ as before.

For obtaining the time evolution operator, we start by finding the eigenvalues and eigenvectors of the pristine system described by $\mathbb{H}_0(\mathbf{k})$. These eigenvalues are readily found by renormalizing one of the bipartite sublattices, since it is
equivalent to consider the squared matrix $\mathbb{H}_{0}^{2}(\mathbf{k})$, as shown in references\cite{Naumis07,Barrios11}. Thus, the eigenvalues of $\mathbb{H}_0(\mathbf{k})$, denoted by $E_{1,2}(k_x,k_y)$ are,
\begin{equation}
\begin{split}
& E_{1}(k_x,k_y)=  \\
 &\pm\sqrt{3+4\cos{(3k_x/2)}\cos{\left(\sqrt{3}k_y/2\right)}+2\cos{(\sqrt{3}k_y)}},\\
 & E_{2}(k_x,k_y)=  \\
 &\pm\sqrt{3-4\cos{(3k_x/2)}\cos{\left(\sqrt{3}k_y/2\right)}+2\cos{(\sqrt{3}k_y)}}.
\end{split}
\end{equation}
To find the unitary transformation that diagonalizes $\mathbb{H}_0(\mathbf{k})$, care must be taken since the eigenvalues of $\mathbb{H}_{0}^{2}(\mathbf{k})$ are degenerate and thus are not necessarily eigenvectors of $\mathbb{H}_0(\mathbf{k})$. However, the eigenfunctions of $\mathbb{H}_0(\mathbf{k})$ correspond to pristine graphene, then one can apply the Bloch theorem for the original Brillouin zone of graphene to get the proper eigenfunctions. Using the well known solution for graphene and ordering the energies as $E_{1}$, $E_{2}$, $-E_{1}$, $-E_{2}$, we obtain the unitary transformation that diagonalizes $\mathbb{H}_0(\mathbf{k})$,
\begin{equation}
O=\left[
\begin{matrix}
\tilde{\mathbb{M}}_a&\tilde{\mathbb{M}}_a\\
\tilde{\mathbb{M}_b}&-\tilde{\mathbb{M}}_b
 \end{matrix}
 \right]
\end{equation}
where
\begin{equation}
\begin{split}
\tilde{\mathbb{M}}_a(\mathbf{k})&=\frac{1}{2}\left[
\begin{matrix}
1&1\\
e^{-i\mathbf{k\cdot\mathbf{a_2}}}&-e^{-i\mathbf{k\cdot\mathbf{a_2}}}
\end{matrix}
\right]\\
\tilde{\mathbb{M}}_b(\mathbf{k})&=\frac{1}{2}\left[
\begin{matrix}
e^{i\theta_{1}}&e^{i\theta_{2}}\\
e^{i\theta_{1}}e^{i\mathbf{k\cdot\mathbf{a_1}}}&-e^{i\theta_{2}}e^{i\mathbf{k\cdot\mathbf{a_1}}}\\
\end{matrix}
\right],
\end{split}
\end{equation}
$\mathbf{a_1}=(3,\sqrt{3})/2$, $\mathbf{a_2}=(3,-\sqrt{3})/2$ being the lattice vectors of pristine graphene, and
\begin{equation}
\begin{split}
e^{i\theta_{1}}&=\frac{1}{E_{1}}\left[1+2e^{-3i k_x/2}\cos{\left(\sqrt{3}k_y/2\right)}\right]\\
e^{i\theta_{2}}&=\frac{1}{E_{2}}\left[1-2e^{-3i k_x/2}\cos{\left(\sqrt{3}k_y/2\right)}\right].
\end{split}
\end{equation}
The next step is to transform Eq. (\ref{uop}) onto the basis in which $\mathbb{H}_0$ is diagonal, {\it i.e.}, to perform $\mathbb{O}^{\dag} \mathbb{U}(\tau) \mathbb{O}=\mathbb{U}^{\prime}(\tau)$. Before doing that, let us reduce $\mathbb{U}(\tau)$ into a simpler form. To do that note that $(\delta\mathbb{H})^2=\left(\delta E (k_y)\right)^2 \,\mathbb{I}_{4\times4}$, where $\mathbb{I}_{4\times4}$ is the $4\times4$ identity matrix and
\begin{equation}
\begin{split}
&\delta E(k_y)=\\
&\sqrt{(\gamma_1-1)^2+(\gamma_2-1)^2+2(\gamma_1-1)(\gamma_2-1)\cos{\left(\sqrt{3}k_y\right)}}.
\end{split}
\label{dE}
\end{equation}

As a result, the exponential of $\delta \mathbb{H}$ can be written as
\begin{equation}
\begin{split}
&\exp{\left[-i\tau(\delta H)\right]}=\cos{\left[\tau\delta E(k_y)\right]}\mathbb{I}-i(\delta H)\frac{\sin{\left[\tau\delta E(k_y)\right]}}{\delta E(k_y)},
\end{split}
\end{equation}
and the time evolution operator becomes,
\begin{equation}
\mathbb{U}(\tau)=\left(\cos{\left[\tau\delta E(k_y)\right]}-i\frac{\sin{\left[\tau\delta E(k_y)\right]}}{\delta E(k_y)}\delta \mathbb{H}(k_y)\right)e^{-i\tau\mathbb{H}_0}.
\end{equation}
Before transforming Eq. (\ref{uop}) into $\mathbb{O}$, we calculate $\delta\mathbb{H}^{\prime}(k_y)=\mathbb{O}^{\dag}\delta\mathbb{H}(k_y) \mathbb{O}$, after some algebraic operations, we get,
\begin{equation}
\delta\mathbb{H}^{\prime}=\left[
\begin{array}{cc}
\delta\tilde{\mathbb{H}}_{+}&\delta\tilde{\mathbb{H}}_{-}\\
-\delta\tilde{\mathbb{H}}_{-}&-\delta\tilde{\mathbb{H}}_{+}
\end{array}
\right],
\end{equation}
where
\begin{equation}
\delta\tilde{\mathbb{H}}_{\pm}=\tilde{\mathbb{M}}^{\dag}_b\,\delta\tilde{\mathbb{H}}\,\tilde{\mathbb{M}}_a\pm\tilde{\mathbb{M}}^{\dag}_a\,\delta\tilde{\mathbb{H}}\,\tilde{\mathbb{M}}_b.
\end{equation}

Finally, the time evolution operator is given by,
\begin{equation}
\begin{split}
&\mathbb{U^{\prime}}(\tau)=-i\frac{\sin{(\tau\delta E)}}{\delta E}\times\\
&\left[
\begin{array}{cc}
\left[\cos{(\tau\delta E)}+\delta\tilde{\mathbb{H}}_{+}\right]\tilde{\mathbb{U}}_{0}(\tau)&\delta\mathbb{H}_{-}\tilde{\mathbb{U}}_{0}^{*}(\tau)\\
-\delta\mathbb{H}_{-}\tilde{\mathbb{U}}_{0}(\tau)&\left[\cos{(\tau\delta E)}-\delta\tilde{\mathbb{H}}_{+}\right]\tilde{\mathbb{U}}_{0}^{*}(\tau)
\end{array}
\right]
\end{split}
\label{Uk}
\end{equation}
where $\mathbb{U^{\prime}}(\tau)=\mathbb{O}^{\dag}(\mathbf{k})\mathbb{U}(\tau)\mathbb{O}(\mathbf{k})$ and
\begin{equation}
\tilde{\mathbb{U}}_{0}(\tau)=
\left[\begin{array}{cc}
e^{i\tau E_1}&0\\
0&e^{i\tau E_2}
\end{array}\right].
\end{equation}
%

\section{Particular case $k_x=0$}
\label{akxcero}

For $k_x=0$ the time evolution operator operator $\mathbb{U}^{\prime}(\tau)$ becomes block diagonal, each block being a $2\times2$ matrix. Hence, the time evolution operator, Eq. (\ref{Uk}), can be written as,
\begin{equation}
\mathbb{U^{\prime}}(\tau)=
\left[\begin{array}{cc}
e^{i\tau \delta\tilde{h}(k_y)}e^{i\tau \tilde{h}_0(k_y)}&0\\
0&e^{-i\tau \delta\tilde{h}(k_y)}e^{-i\tau \tilde{h}_0(k_y)}
\end{array}\right]
\label{Ukx0}
\end{equation}
where
\begin{equation}
\begin{split}
\delta\tilde{h}(k_y)&=\delta h(k_y)\,\delta\hat{\mathbf{h}}\cdot\mathbf{\sigma},\\
\tilde{h}_0(k_y)&=\mathbb{I}_{2\times2}+2\cos{\left(\sqrt{3}k_y/2\right)}\sigma_z,
\end{split}
\end{equation}
with $\mathbf{\sigma}=(\sigma_x,\sigma_y,\sigma_z)$, $\sigma_i$ ($i=x,y,z$) being the $2\times2$ Pauli matrices written in the basis at where $\sigma_z$ is diagonal, $\mathbb{I}_{2\times2}$ the $2\times2$ identity matrix, and the components of $\delta\mathbf{h}$ are
\begin{equation}
\begin{split}
\delta h^{(y)}&=(\gamma_1-\gamma_2)\sin{\left(\sqrt{3}k_y/2\right)},\\
\delta h^{(z)}&=(\gamma_1+\gamma_2-2)\cos{\left(\sqrt{3}k_y/2\right)}.
\end{split}
\end{equation}
we have also defined the norm $\delta h(k_y)=|\delta\mathbf{h}|$.

By using the addition rule of SU(2), Eq. (\ref{Ukx0}) can be written as follows,
\begin{equation}
\mathbb{U^{\prime}}(\tau)=
\left[\begin{array}{cc}
e^{i\tau \tilde{h}_{\text{eff}}(k_y)}&0\\
0&e^{-i\tau \tilde{h}_{\text{eff}}(k_y)}
\end{array}\right]
\end{equation}
where
\begin{equation}
\tilde{h}_{\text{eff}}(k_y)=\mathbb{I}_{2\times2}+\Omega(k_y)\hat{\mathbf{h}}_{\text{eff}}\cdot\mathbf{\sigma},
\end{equation}
where $\Omega(k_y)$ is given by
\begin{equation}
\begin{split}
&\cos{\left[\tau\Omega(k_y)\right]}=\cos{(\tau\delta h)}\cos{\left[2\tau\cos{\left(\sqrt{3}k_y/2\right)}\right]}\\
&-\frac{\delta h^{(z)}}{\delta h}\sin{(\tau\delta h)}\sin{\left[2\tau\cos{\left(\sqrt{3}k_y/2\right)}\right]}.
\end{split}
\end{equation}
Finally, the unit vector $\hat{\mathbf{h}}_{\text{eff}}$ is 
\begin{equation}
\begin{split}
\hat{\mathbf{h}}_{\text{eff}}=&\frac{1}{\sin{\left[\tau\Omega(k_y)\right]}}\left(\delta \hat{\mathbf{h}}\sin{(\tau\delta h)}\cos{\left[2\tau\cos{\left(\sqrt{3}k_y/2\right)}\right]}\right. \\
&\left.+\hat{e}_z\cos{(\tau\delta h)}\sin{\left[2\tau\cos{\left(\sqrt{3}k_y/2\right)}\right]}\right. \\
&\left.-\delta h^{(y)}\hat{e}_x\sin{(\tau\delta h)}\sin{\left[2\tau\cos{\left(\sqrt{3}k_y/2\right)}\right]}\right).
\end{split}
\end{equation}

It is noteworthy that the quasienergies of the system are thus given by $\pm\tau\omega(k_y)=\pm\tau(\pm1+\Omega(k_y))$, see the main text.

\bibliography{biblioArmChairGraphene}{}
\end{document}